\documentclass[aps,pra,twocolumn,superscriptaddress,groupedaddress]{revtex4}

\usepackage{amssymb} \usepackage{color,graphicx} \usepackage{amsmath}
\usepackage{amsbsy} \usepackage{amsthm} \usepackage{bbm}
\usepackage{bm,bbm} \usepackage{float} \usepackage{braket}
\usepackage{placeins}
\usepackage[colorlinks=true,citecolor=blue,linkcolor=red,urlcolor=red]{hyperref}
\usepackage[sort&compress]{natbib}
\usepackage{comment}
\usepackage{stackengine}

\newcommand{\bq}{\begin{equation}} \newcommand{\eq}{\end{equation}}
\newcommand{\bqali}{\bq\begin{aligned}}
\newcommand{\eqali}{\end{aligned}\eq}
\newcommand{\bqn}{\begin{equation*}}
\newcommand{\eqn}{\end{equation*}}

\newcommand\D{\operatorname{d}\!}

\renewcommand\r{{\bf r}}
\newcommand\p{{\bf p}}

\newcommand\kb{k_\text{\tiny B}}

\newcommand\com[2]{[#1,#2]}
\newcommand\acom[2]{\{#1,#2\}}

\newcommand\DNS{\mathcal S_{xx}(\omega)}
\newcommand\DNSL[1]{\mathcal S^{#1}_\text{\tiny L}(\omega)}

\graphicspath{{../Figures/}}

\begin{document}

\title{Testing the gravitational field generated by a quantum superposition}

\author{M. Carlesso}
\email{matteo.carlesso@ts.infn.it}
\affiliation{Department of Physics, University of Trieste, Strada Costiera 11, 34151 Trieste, Italy}
\affiliation{Istituto
Nazionale di Fisica Nucleare, Trieste Section, Via Valerio 2, 34127 Trieste,
Italy}

\author{A. Bassi}
\affiliation{Department of Physics, University of Trieste, Strada Costiera 11, 34151 Trieste, Italy}
\affiliation{Istituto
Nazionale di Fisica Nucleare, Trieste Section, Via Valerio 2, 34127 Trieste,
Italy}

\author{M. Paternostro}
\affiliation{Centre for Theoretical Atomic, Molecular and Optical Physics, School of Mathematics and Physics, Queen\textquoteright{}s University, Belfast BT7 1NN, United Kingdom}

\author{H. Ulbricht}
\affiliation{School of Physics and Astronomy, University of Southampton, Southampton SO17 1BJ, United Kingdom}

\date{\today}

\begin{abstract}
{What gravitational field is generated by a massive quantum system in a spatial superposition? Despite decades of intensive theoretical and experimental research, we still do not know the answer. On the experimental side, the difficulty lies in the fact that gravity is weak and requires large masses to be detectable. However, it becomes increasingly difficult to  generate spatial quantum superpositions for increasingly large masses, in light of the stronger environmental effects on such systems. Clearly, a delicate balance between the need for strong gravitational effects and weak decoherence should be found. We show that such a trade off could be achieved in an optomechanics scenario that allows to determine whether the gravitational field generated by a quantum system in a spatial superposition is in a coherent superposition or not. We estimate the magnitude of the effect and show that it offers perspectives for observability.}
\end{abstract}

\pacs{xxxxxxxxxxxx}

\maketitle


{Quantum field theory is one of the most successful theories ever formulated. All matter fields, together with the electromagnetic and nuclear forces, have been successfully  embedded in the quantum framework. They form the standard model of elementary particles, which not only has been confirmed in all advanced accelerator facilities, but has also become an essential ingredient for the description of the universe and its evolution. }

{In light of this, it is natural to seek a quantum formulation of gravity as well. Yet, the straightforward procedure for promoting the classical field as described by general relativity, into a quantum field, does not work. Several strategies have been put forward, which turned into very sophisticated theories of gravity, the most advanced being string theory and loop quantum gravity. Yet,  none of them has reached the goal of providing a fully consistent quantum theory of gravity.  }

{At this point, one might wonder whether the very idea of quantizing gravity is correct~\cite{Penrose:2014aa, Dyson:2013aa, Adler:2015aa, Feynman:1995aa, Penrose:1996aa,Penrose:1998aa,Carlip:2008aa,Kibble:1981aa,Mattingly:2005aa,Kiefer:2007aa,Moller:1962aa,Rosenfeld:1963aa,Eppley:1977aa,Page:1981aa,Mattingly:2006aa,Albers:2008aa,Bronstein:2012aa}. At the end of the day, according to general relativity, gravity is rather different from all other forces. Actually, it is not a force at all, but a manifestation of the curvature of spacetime, and there is no obvious reason why the standard approach to the quantization of fields should work for spacetime as well.  A future unified theory of quantum and gravitational phenomena might require a radical revision not only of our notions of space and time, but also of (quantum) matter. This scenario is growing in likeliness. }

{From the experimental point of view, it has now been ascertained that quantum matter (i.e.~matter in a genuine quantum state, such as a coherent superposition state) couples to the Earth's gravity in the most obvious way. This has been confirmed in neutron~\cite{Colella:1975aa}, atom~\cite{Peters:1999aa} interferometers and used for velocity selection in molecular interferometry~\cite{Brezger:2002aa}. However, in all cases, the gravitational field is classical, i.e.~it is generated by a distribution of matter (the Earth) in a fully classical state.  Therefore, the plethora of successful experiments mentioned above does not provide hints, unfortunately, on whether gravity is quantum or not. }

{The large attention and media coverage about the BICEP2 Collaboration's experiment having shown the quantum origin of primordial gravitational fluctuations~\cite{Ade:2014aa}, subsequently disproved by Planck Collaboration's data analysis~\cite{Planck-Collaboration:2016aa}, testifies the importance and urgency of a pragmatic assessment of the question of whether gravity is quantum or not.}

{In this paper, we discuss an approach where a quantum system is forced in the superposition of two different positions in space, and its gravitational field is explored by a probe (Fig.~\ref{fig2}). Using the exquisite potential for transduction offered by optomechanics, we can in principle determine whether the gravitational field is the superposition of the two gravitational fields associated to the two different states of the system, or not. The first case amounts to a quantum behavior of gravity, the second to a classical-like one. }

The remainder of this manuscript is organized as follows. In Sec.~\ref{frame} we define the context considered throughout the manuscript and discuss both the quantum and semi-classical scenarios for gravity. Sec.~\ref{modello} presents the theoretical model for the dynamics of the optomechanical platform that we address, while Sec.~\ref{reveal} puts forward our proposals for the inference of the difference between a quantum and classical nature of  gravity. Finally, in Sec.~\ref{conclusioni} we state our conclusions and discuss a few interesting features of our findings.

\section{Framework} 
\label{frame}
\begin{figure}[t!]
\hskip0.17\columnwidth{\bf (a)}\hskip0.4\columnwidth{\bf (b)}
\includegraphics[width=0.6\columnwidth]{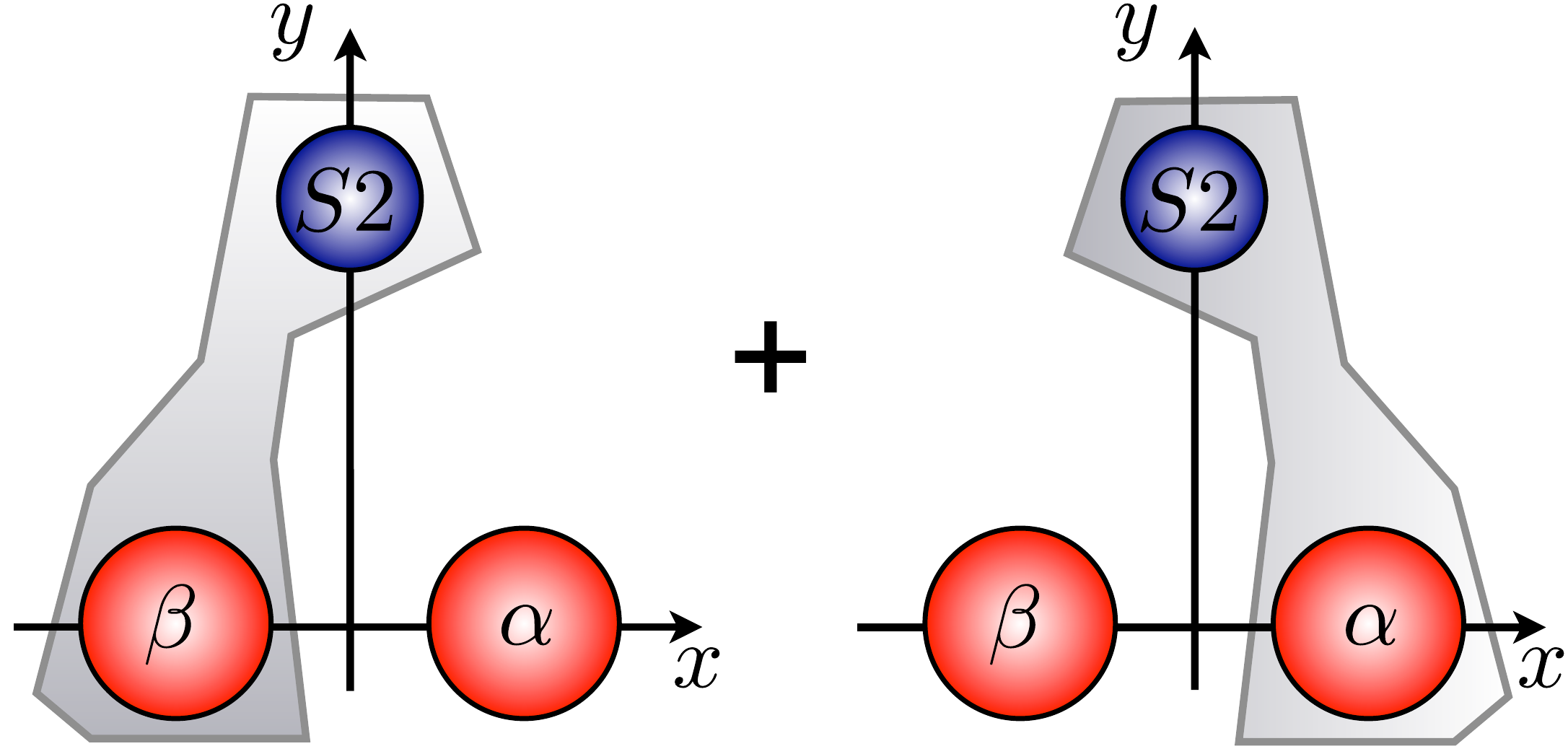}
\includegraphics[width=0.28\columnwidth]{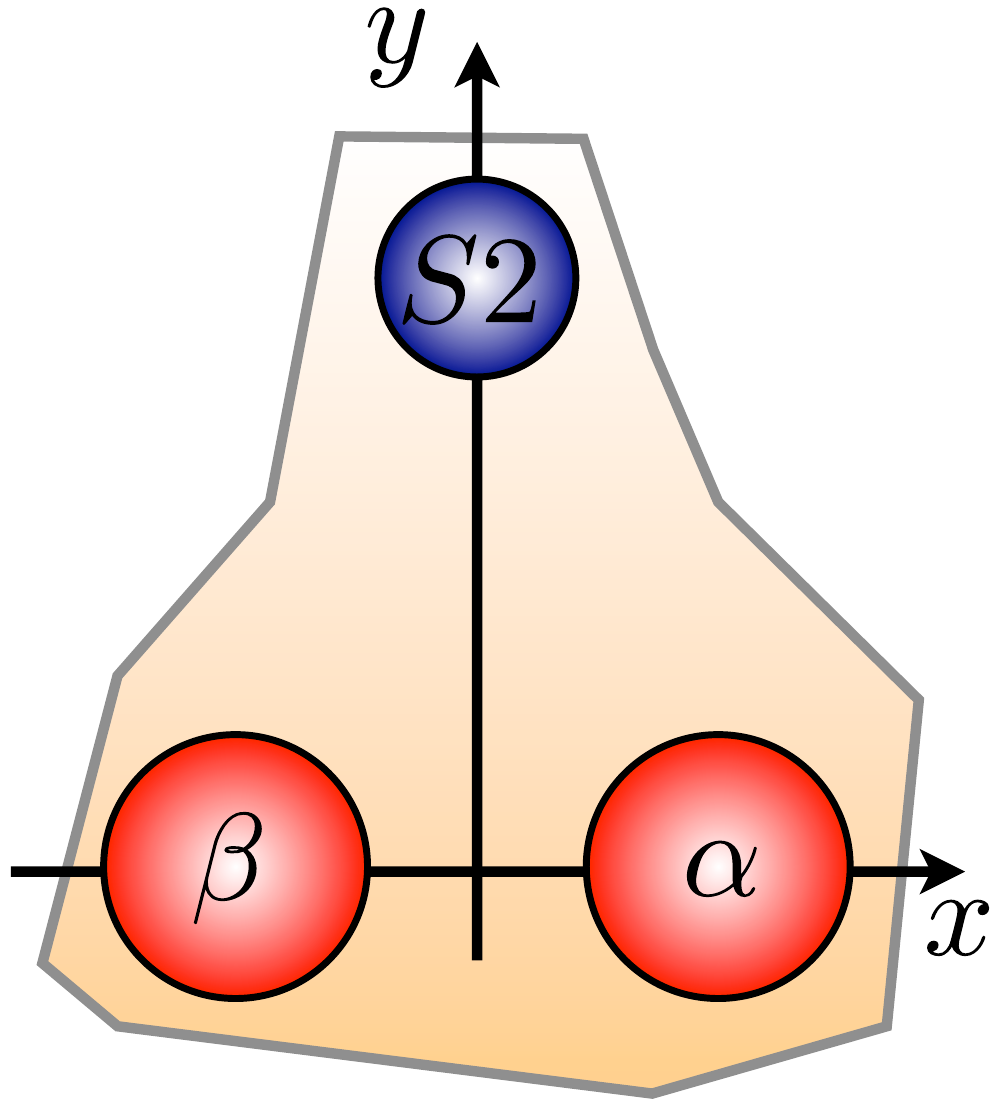}
\caption{Schematic representation of the two-body setup. S1 is prepared in a spatial superposition along the $x$ direction (red balls). S2 is initially prepared in a localized wavepacket (blue ball), and it probes the gravitational field generated by S1. ({\bf a}) The gravitational field acting on S2 is a linear combination of gravitational fields produced by S1 being in a superposed state. ({\bf b}) The semi-classical treatment of gravity, where the gravitational field acting on S2 is that produced by a total mass $m_1$ with density $\frac12\left(|\alpha(\r)|^2+|\beta(\r)|^2\right)$.
}
\label{fig2}
\end{figure}

We consider a setup formed of two systems interacting gravitationally. All non-gravitational interactions are considered, for all practical purposes, negligible. The first system (S1) has a mass $m_1$, and it is initially prepared in a spatial superposition along the $x$ direction. Its wave-function is $\psi(\r_1)=\tfrac{1}{\sqrt{2}}(\alpha(\r_1)+\beta(\r_1))$, where $\alpha(\r_1)$ and $\beta(\r_1)$ are sufficiently well localized states in position, far from each other in order to prevent any overlap. Thus, we can consider them as distinguishable (in a macroscopic sense), and we approximate $\braket{\alpha|\beta}\simeq0$. The second system (S2) will serve as a probe of the gravitational field generated by S1, it has mass $m_2$ and state $\phi(\r_2)$. The state $\phi(\r_2)$ is initially assumed to be localized in position and centered along the $y$ direction [cf.~Fig.~\ref{fig2}].
The question we address is: which is the gravitational field, generated by the quantum superposition of S1, that S2 experiences? We probe the following two different scenarios.

\noindent {\it Quantum gravity scenario}.--
Although we do not have a quantum theory of gravity so far, one can safely claim that it would manifest in S1 generating a superposition of gravitational fields. The linearity, which is the characteristic trait of quantum theory, is preserved, as one would expect in any quantum theory of gravity. The reaction of S2 is then to go in a superposition of being attracted towards the region where $\ket\alpha$ sits and where $\ket\beta$ does. The final two-body state will have the following entangled form
\bq\label{final.ent.2}
\Psi^\text{\tiny final}_\text{\tiny QG}(\r_1,\r_2)=\frac{\alpha({\bf r}_1)\phi_\alpha({\bf r}_2)+\beta({\bf r}_1)\phi_\beta({\bf r}_2)}{\sqrt{2}},
\eq
where $\phi_\alpha({\bf r}_2)$ ($\phi_\beta({\bf r}_2)$) represents the state of S2 attracted towards the region where $\ket\alpha$ ($\ket\beta$) rests. 
The latter superposition of motions for S2 is produced by the following potential
\bq
\label{V-qm}
\hat  V_\gamma (\hat \r_2) = -G m_1 m_2\,\int\D{\r}_1\frac{|\gamma({ \r}_1)|^2}{| {\r}_1-\hat { \r}_2|},~~~~~(\gamma=\alpha,\beta).
\eq
Moreover, we assume that the quantum fluctuations around the mean values for S1 are small, so that the gravitational interaction can be approximated by
\begin{equation}\label{V-qm-approx}
\hat  V_{\gamma} (\hat \r_2) \approx -\frac{G m_1 m_2}{|\braket{\hat {\bf r}_1(t)}_{\gamma}-\hat{\bf r}_2(t)|},~~~~~~(\gamma=\alpha,\beta),
\end{equation}
where $\braket{\hat\r_1}_\gamma=\braket{\gamma|\hat\r_1|\gamma}$ with $\gamma=\alpha, \beta$.

\noindent {\it Semiclassical gravity scenario}.--
The second scenario sees gravity as fundamentally classical. In this case, it is not clear which characteristics one should expect from the gravitational field generated by a superposition. However, in analogy with classical mechanics, one can assume that is the mass density of the system in superposition that produces the gravitational field. This is also what is predicted by the Schr\"odinger-Newton equation \cite{Bahrami:2014ab,Diosi:1984aa,Penrose:1996aa,Giulini11,Giulini12,Giulini13}. In such a case, what matters is the full wavefunction of S1 and not its single parts. Consequently, the generated gravitational field is not in a quantum superposition, but it manifests as that produced by a classical object with total mass $m_1$ with density $|\psi(\r_1)|^2\simeq\frac12\left(|\alpha(\r_1)|^2+|\beta(\r_1)|^2\right)$. Clearly, S2 reacts as driven by a classical gravitational field.
The final two-body state will be of the form
\bq
\Psi^\text{\tiny final}_\text{\tiny CG}(\r_1,\r_2)=\frac{\alpha({\bf r}_1)+\beta({\bf r}_1)}{\sqrt{2}}\phi(\r_2),
\eq
where the difference with Eq.~\eqref{final.ent.2} is clear. The gravitational potential becomes
\bq
\hat  V_\text{\tiny cl} (\hat \r_2) \approx \tfrac12\sum_{\gamma=\alpha,\beta} \hat  V_{\gamma} (\hat \r_2),
\eq
where $\hat  V_{\gamma} (\hat \r_2)$ can be eventually approximated as in Eq.~\eqref{V-qm-approx}.

In the next Section, we investigate the difference between the two scenarios by exploiting the sophisticated and powerful machinery provided by optomechanics. 

\begin{figure}[t!]
\centering
\includegraphics[width=0.65\columnwidth]{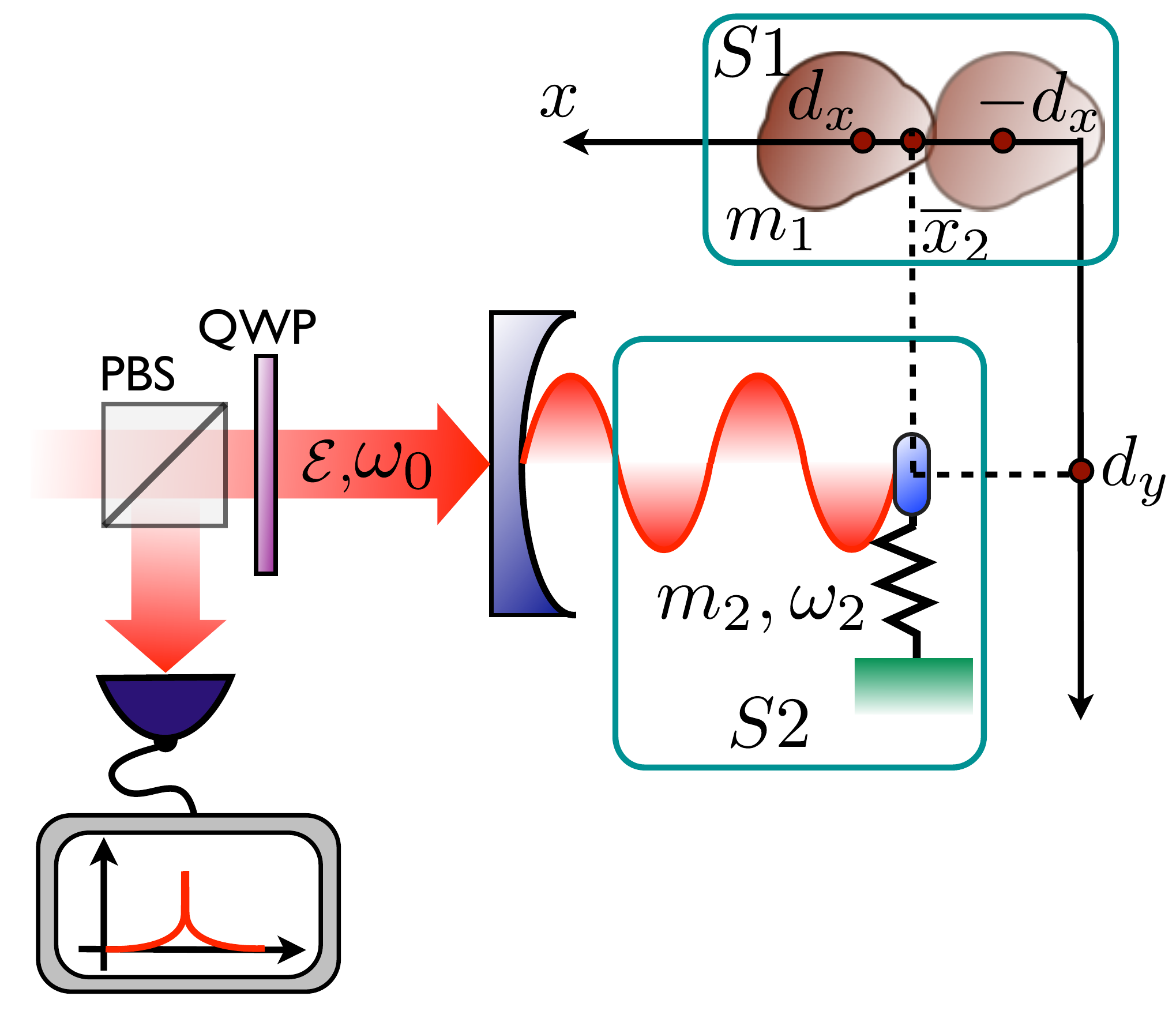}
\caption{The proposed set-up for the optomechanical falsification of quantum/classical gravity. A system S1 is prepared in a superposition of two localised states at $\pm d_x$ along the $x$ axis. An optomechanical cavity acts as transducer and probe of (potentially quantum) gravity effects S2: the effect of the gravitational coupling between S1 and the mechanical oscillator of an optomechanical cavity induces an effect on the variance of the position fluctuations of the oscillator. The mean position of the latter along the $x$ axis is $\bar x_2$. The cavity is pumped by an external field (frequency $\omega_0$ and coupling rate ${\cal E}$).}
\label{fig3}
\end{figure}

 \section{Theoretical model}
 \label{modello}
To describe the dynamics that follow the first or second scenario, we take advantage of the quantum Langevin equations, which is the typical description for optomechanical systems. Moreover, we assume that the mass of S1 is sufficiently large to consider an adiabatic approach: S1 is stable and S2 evolves in the gravitational potential produced by S1. Clearly, such a situation can be assumed only as long as the S1 superposition lives.
We assume S2 as trapped harmonically in $\r_\text{\tiny osc}=(r_{x,\text{\tiny osc}},r_{y,\text{\tiny osc}},0)$ along the $x$ and $y$ directons by means of the cavity fields.
The corresponding quantum Langevin equations for the position ${\hat r}_i$ and momentum $\hat p_i$ operator of S2 read \cite{Aspelmeyer}
\begin{equation}
\label{differ_eqs}
\begin{aligned}
{\frac{\D{\hat r}_i(t)}{\D t} }&=\dfrac{{\hat p}_i(t)}{m_2},\\
{\frac{\D{\hat p}_i(t)}{\D t}  }&= -m_2\omega_i^2 \left({\hat r}_i(t) -r_{i,\text{\tiny osc}}\right)
-\gamma_i {\hat p}_i(t)
+{\hat \xi}_i(t)\\
&+\hbar \chi_i\hat a_i^\dag(t)\hat a_i(t)+\dfrac{i}{\hbar}\com{\hat V_{{\nu}}}{\hat p_i(t)},
\end{aligned}
\end{equation}
where $i=x,y$ (we do not consider the motion along $z$) and $\nu=\alpha, \beta, \text{cl}$. Here, $\omega_i$ is the harmonic frequency of the mechanical oscillator,
$\gamma_i$ is the damping rate for the vibrations, which are characterized by the noise operator $\hat \xi_i$, having the correlation functions defined as $\braket{\hat \xi_i(t)}=0$ and
\bq
\braket{\hat \xi_i(t)\hat \xi_j(s)}=\hbar m\gamma_i\delta_{ij}\!\int\!\frac{\D\omega}{2\pi}e^{-i\omega(t-s)}\omega[1+\coth(\tfrac{\hbar\omega}{2\kb T})].
\eq
The position of S2 is measured by means of the cavity field, whose creation and annihilation operator are $\hat a_i^\dag$ and $\hat a_i$. The dynamical equation of the latter is given by 
\bq
{\frac{\D {\hat a_i}(t)}{\D t}   }= -i\left[\Delta_{0,i}- \chi_i{\hat r}_i(t)\right] {\hat a_i}(t)-\kappa_i\hat a_i(t)
+\sqrt{2\kappa_i} {\hat a}_{i,\text{\tiny in}}(t),\\
\eq
where we defined $\Delta_{0,i}=\omega_{c,i}-\omega_{0,i}$, with $\omega_{0,i}$ denoting the frequency of the external laser,
$\omega_{c,i}$ the frequency of the cavity mode derived by the laser,
$\chi_i=\omega_{c,i}/L_i$  the optomechanical coupling constant between the cavity and the mechanical oscillator with $L_i$ the size of the cavity,
and ${\cal E}_i=\sqrt{2\kappa_i{\cal P}_i/\hbar \omega_{0,i}}$. Here, ${\cal P}_i$ is the laser power and $\kappa_i$ is the cavity photon decay rate.
Moreover, we defined ${\hat a}_{i,\text{\tiny in}}$ as the annihilation operator of external laser field, whose only non-zero correlation reads $\braket{\hat a_{i,\text{\tiny in}}(t)\hat a^\dag_{j,\text{\tiny in}}(s)}=\delta_{ij}\delta(t-s)$. The last term in Eq.~\eqref{differ_eqs} describes the gravitational interaction with S1, whose action is described below.

To be quantitative, we define the mean positions of the two systems in interaction. We consider S1 as holding a steady position that can be approximated to its average value on $\alpha$ or $\beta$ respectively: {$\braket{\hat  \r_1(t)}_{\gamma}\approx(s_\gamma d_x,0,0)$, with $s_\alpha=1$, $s_\beta=-1$. Conversely, we consider the position of S2 as an operator, center in $(\bar x_2,d_y,0)$ [cf.~Fig.~\ref{fig2}]. Thus, we have $\hat \r_2(t)=(\hat r_x(t),\hat r_y(t), 0)=(\bar x_2+\hat \delta_x(t),d_y+\hat \delta_y(t),0)$ and $\hat\p_2(t)=(\hat p_x(t), \hat p_y(t),0)$ is its momentum operator.

Assuming that the quantum fluctuations $\delta \hat {\bf r}_2(t)=(\hat \delta_x(t),\hat \delta_y(t),0)$ around the {initial} mean values for S2 are small, we can expand the commutator in the last term of Eq.~\eqref{differ_eqs} up to the first order in the fluctuations. Thus, we have
\bq\label{defcomVx}
\frac{i}{\hbar}\com{\hat V_{{\nu}}}{\hat p_i(t)}=C^{(\nu)}_{0,i}+C^{(\nu)}_{1,i}\hat\delta_i(t)+C^{(\nu)}_{2,i}\hat\delta_j(t),~~\text{with }j\neq i.
\eq
In the quantum scenario, the coefficients $C^{(\nu)}_{n,i}$ entering in Eq.~\eqref{defcomVx} are defined in Table~\ref{tavola}, while those in the classical scenario are given by $C^\text{(cl)}_{n,i}=\tfrac12(C^{(\alpha)}_{n,i}+C^{(\beta)}_{n,i})$. 
\begin{table}[t!]
\begin{tabular}{|c|c|c|}
\hline\hline
 \multicolumn{3}{|c|}{Quantum scenario}\\
 \hline
$C^{(\gamma)}_{n,i}$&$i=x$& $i=y$ \\ 
 \hline
$n=0$&$\mathcal G_\gamma(s_\gamma d_x-\bar x_2)$&$\mathcal G_\gamma d_y$\\ 
$n=1$&$\frac{\mathcal G_\gamma}{h_\gamma^2}\left[3(\bar x_2-s_\gamma d_x)^2-h_\gamma^2\right]$&$\frac{\mathcal G_\gamma}{h_\gamma^2}(3d_y^2-h_\gamma^2)$\\ 
$n=2$ &$-\frac{3 \mathcal G_\gamma}{h_\gamma^2}(s_\gamma d_x-\bar x_2)d_y$ &$-\frac{3 \mathcal G_\gamma}{h_\gamma^2}(s_\gamma d_x-\bar x_2)d_y$\\ 
\hline\hline
\end{tabular}
\caption{\label{tavola}Explicit form of the coefficients $C^{(\gamma)}_{n,i}$ entering in Eq.~\eqref{defcomVx} for the quantum scenario, with $\mathcal G_\gamma={G m_1m_2}/{h_\gamma^3}$ and $h_\gamma=\sqrt{(\bar x_2-s_\gamma d_x)^2+d_y^2}$. For the classical scenario we have $C^\text{(cl)}_{i,x}=\tfrac12(C^{(\alpha)}_{i,x}+C^{(\beta)}_{i,x})$. }
\end{table}
 In the limit of $d_x\gg\bar x_2$, they become
\begin{subequations}\label{defCy}
\begin{gather}
C_{1,x}=\dfrac{G m_1m_2}{d^5}(2d_x^2-d_y^2),\\
C_{1,y}=\dfrac{G m_1m_2}{d^5}(2d_y^2-d_x^2),\\
C_{2}^{(\gamma)}=-\frac{3Gm_1m_2}{d^5}d_xd_ys_\gamma,~~\text{and}~~C_{2}^{\text(cl)}=0,\label{eq.C2}
\end{gather}
\end{subequations}
where $d^2=({d_x^2+d_y^2})$.
Here only $C_{2}^{(\nu)}$ depends on the specific scenario (quantum or semi-classical) we are considering. Following conventional approach, one finds: 
\bq
\bar r_i^{(\nu)}=\frac{\hbar \chi_i|\bar a_i|^2+C_{0,i}^{\nu}}{m_2\omega_i^2}+r_{i,\text{\tiny osc}},\quad\text{and}\quad\bar p^{(\nu)}_i=0.
\eq
We can remove the radiation pressure contribution by setting the center of the harmonic trap to $r_{i,\text{\tiny osc}}=-\hbar \chi_i|\bar a_i|^2/m_2\omega_i^2$. Moreover, we assume that $d_x\gg\bar x_2$, such that one can approximate $h_\gamma\simeq d=(d_x^2+d_y^2)^{1/2}$ [cf.~Table~\ref{tavola}], thus finding
\bq
\bar x_2^{(\gamma)}=\frac{Gm_1d_x}{\omega_x^2d^3}s_\gamma,~~~\bar y_2^{(\gamma)}=\frac{Gm_1d_y}{\omega^2_yd^3}.
\eq
These expressions show the first difference between the quantum and the classical scenario. In the quantum scenario S2 is pulled towards positive (or negative) $x$ while in the classical scenario it remains at the center $\bar x_2^{\text(cl)}=\bar x_2^{(\alpha)}+\bar x_2^{(\beta)}=0$. However, it also highlights the difficulties one has in discerning the two scenarios. Once the average is taken in the quantum scenario, we have $\braket{\hat x_2}_\text{(qu)}=\tfrac12\sum_\gamma\bar x_2^{(\gamma)}=0$, which corresponds to the classical result.

Equation \eqref{defcomVx} shows that the difference between the quantum and the semi-classical scenario is embedded in the coupling between the motions along $x$ and $y$ of S2. Indeed, in the quantum scenario, the gravitation attraction of S1 pulls S2 towards one of the branches of the superposition of S1, leading to correlations between the $x$ and $y$ motions.
Conversely, in the semi-classical scenario, for which $C_2^{\text (cl)}=0$, the dynamics along the two direction is decoupled, due to the symmetrical  attraction of S1 along $y$.
The verification of a coupling of the motion along $x$ with that along $y$ would be sufficient to prefer the quantum scenario over the semi-classical one. Next we discuss possible mechanisms that can be exploited for this task.

\section{Revelation strategies}
\label{reveal}
There are different measurements that one can exploit for witnessing the correlations between the $x$ and $y$ motions, and thus providing a verification of the quantum scenario over the semi-classical one.

{\it 1) Direct measurement of the Density Noise Spectrum.} To quantify the difference between the two scenarios, we consider the Density Noise Spectrum (DNS) corresponding to the motion of S2 along the $x$ axis. By working under conditions such that $d_x\gg\bar x_2$, the Langevin equations for the fluctuations read
\begin{equation}\label{eqdeltap}
\begin{aligned}
{\frac{\D{\hat\delta_i}(t)}{\D t} }&={\frac{{\delta\hat p}_i(t)}{m_2},}\\
{\frac{\D {\delta\hat p}_i(t)}{\D t}   }&= -m_2\omega_i^2\hat\delta_i(t)-\gamma_i {\delta\hat p}_i(t){+{\hat \xi}_i(t)}+C_{1,i} \hat \delta_i(t)\\ 
&+C_{2,i}^{(\nu)} \hat \delta_j(t)+\hbar\chi_i[\bar a_i^*\delta\hat a_i(t)+\bar a_i\delta a_i^\dag(t)],\\
{\frac{\D {\delta\hat a_i}(t)}{\D t}   }&{=} -i\Delta_{i}^{(\nu)} {\delta\hat a_i}(t){+}i \chi_i\bar a_i{\hat \delta_i}(t){-}\kappa_i\delta\hat a_i(t){+}\sqrt{2\kappa_i} {\hat a}_{i,{\text{\tiny in}}}(t) 
\end{aligned}
\end{equation}
for $j\neq i$. The coefficients $C_{n,i}^{(\nu)}$ are approximated as in Eqs.~\eqref{defCy}, $\Delta_i^{(\nu)}=\Delta_{0,i}-\chi_i\bar a_i\bar r_i^{(\nu)}$, which becomes $\Delta_i\simeq\Delta_{0,i}$ in light of the weakness of the optomechanical coupling. 

Eqs.~\eqref{eqdeltap} can be solved in the frequency domain by using the standard approach \cite{Aspelmeyer}. By defining $\tilde r_i(\omega)$ as the Fourier transform of $\hat\delta_i(t)$, after lengthly yet straightforward calculations, we find
\begin{widetext}
\bq\label{solxy}
\tilde r_i(\omega)=\frac{1}{m_2\left[\omega_{i,\text{\tiny eff}}^2(\omega)-\omega^2-i\gamma_{i,\text{\tiny eff}}(\omega)\omega\right]}\left[\tilde\xi_i(\omega)+C_2^{(\nu)}\tilde r_j(\omega)+\hbar \chi_i\sqrt{2\kappa_i}\left(\frac{\bar a_i^*\tilde a_{i,\text{\tiny in}}(\omega)}{\kappa_i+i(\Delta_i-\omega)}	+\frac{\bar a_i\tilde a^\dag_{i,\text{\tiny in}}(\omega)}{\kappa_i-i(\Delta_i+\omega)}	\right)	\right],
\eq
\end{widetext}
where we defined the following effective frequencies and dampings
\begin{subequations}\label{omegagammaeff}
\bq
\omega_{i,\text{\tiny eff}}^2(\omega)=\omega_i^2+\frac{2\hbar\chi_i^2|\bar a_i|^2\Delta_i(\omega^2-\kappa_i^2-\Delta_i^2)}{m_2\left[(\kappa_i^2+\Delta_i^2+\omega^2)^2-4\Delta_i^2\omega^2\right]}-\frac{C_{1,i}}{m_2},
\eq
\bq
\gamma_{i,\text{\tiny eff}}(\omega)=\gamma_i+\frac{4\hbar\chi_i^2|\bar a_i|^2\Delta_i\kappa_i}{m_2\left[(\kappa_i^2+\Delta_i^2+\omega^2)^2-4\Delta_i^2\omega^2\right]}.
\eq
\end{subequations}
The effect of such correlation can be seen in the DNS, which can be derived from Eq.~\eqref{solxy} by applying its definition $\mathcal S_{ii}(\omega)=\tfrac{1}{4\pi}\int\D\Omega\,\braket{\acom{\tilde r_i(\omega)}{\tilde r_i(\Omega)}}$. Then we find
\begin{widetext}
\bq\label{eqDNS}
\DNS=\frac{m_2g_y(\omega)\left[\left(\hbar	m_2\gamma_x\omega\coth\left(\tfrac{\hbar\omega}{2\kb T}\right)+\DNSL{x}\right)+\frac{(C_2^{(\nu)})^2}{m_2^2g_y^2(\omega)}\left(\hbar	m_2\gamma_y\omega\coth\left(\tfrac{\hbar\omega}{2\kb T}\right)+\DNSL{y}\right)	\right]}{m_2^4g_x(\omega)g_y(\omega)-2m_2^2(C_2^{(\nu)})^2f(\omega)+(C_2^{(\nu)})^4},
\eq
\end{widetext}
where
\begin{subequations}
\begin{gather}
g_i(\omega)=(\omega_{i,\text{\tiny eff}}^2(\omega)-\omega^2)^2+\gamma_{i,\text{\tiny eff}}^2(\omega)\omega^2,\\
\DNSL{i}=\frac{2\hbar^2\chi_i^2\kappa_i|\bar a_i|^2(\kappa_i^2+\Delta_i^2+\omega^2)}{\left[(\kappa_i^2+\Delta_i^2+\omega^2)^2-4\Delta_i^2\omega^2\right]},
\end{gather}
and
\bqali
f(\omega)&=(\omega_{x,\text{\tiny eff}}^2(\omega){-}\omega^2)(\omega_{y,\text{\tiny eff}}^2(\omega){-}\omega^2){-}\gamma_{x,\text{\tiny eff}}(\omega)\gamma_{y,\text{\tiny eff}}(\omega)\omega^2.
\eqali
\end{subequations}
with $\omega_\text{\tiny eff}$ and $\gamma_\text{\tiny eff}$ denoting the effective frequency and damping respectively. Eq.~\eqref{eqDNS} shows that in the quantum scenario the gravitational interaction leads to an extra contribution in the DNS (last term in squared brackets), which is directly connected to the motion along $y$.
Such a term appears as an extra peak centred in the effective oscillation frequency of the $y$ motion. The amplitude of the peak is related to the coupling between S2 and the cavity field along $y$. Clearly, the larger the coupling the bigger is the amplitude of the peak. An example of the presence of this second peak is shown in Fig.~\ref{fig:piccoy}.

\begin{figure}[t!]
\centering
\includegraphics[width=\linewidth]{{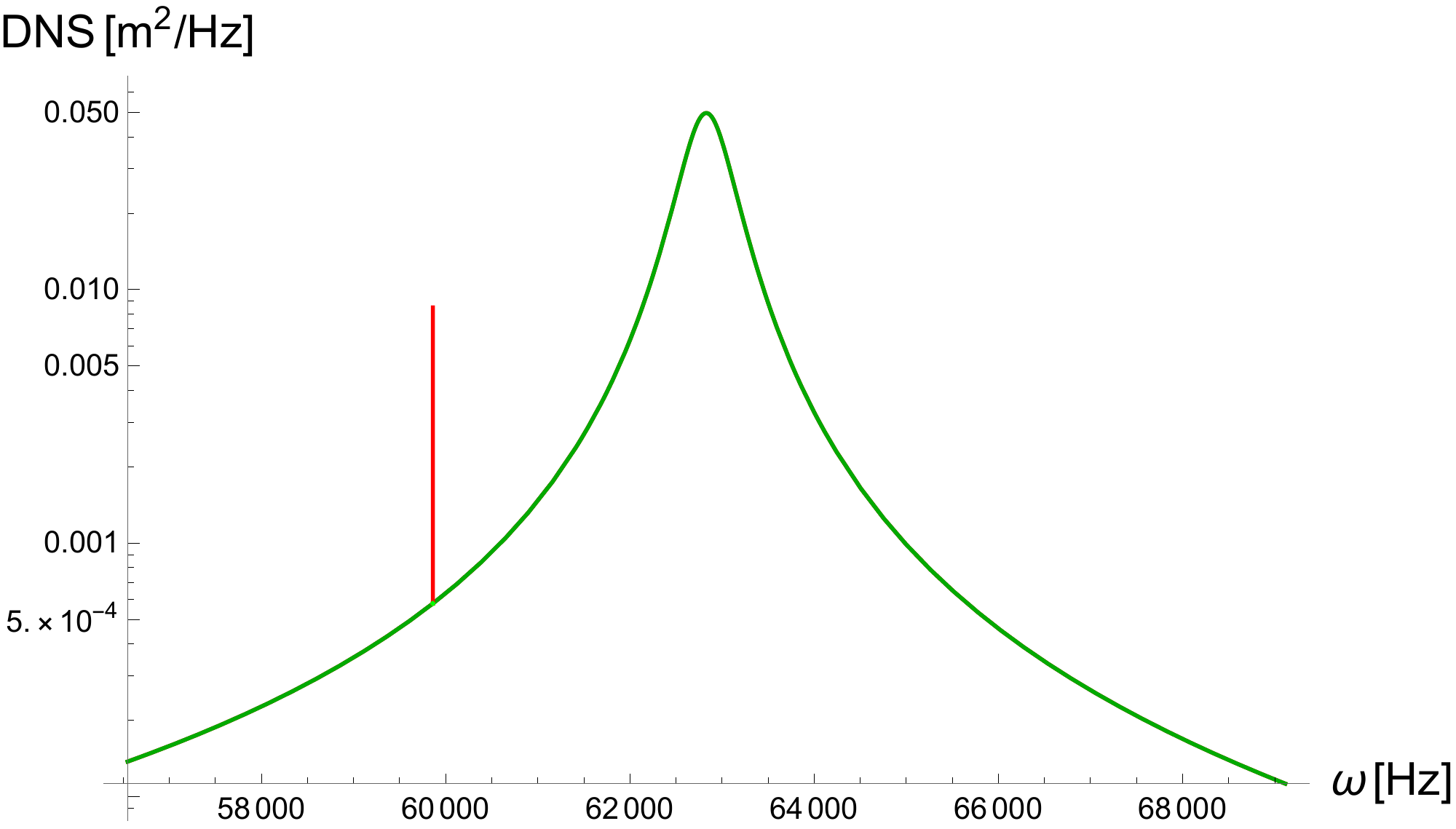}}
\caption{\label{fig:piccoy} Comparison between the DNS for the classical (in green) and the quantum (in red) scenario. We have taken $m_1=5\times 10^{-14}\,$kg, $m_2=9.5\times10^{-19}$\,kg, $d_{x}=10^{-9}$\,m, $d_{y}=2.9 \times10^{-4}$\,m, $\omega_x=2\pi\times10^4$\,Hz, $\omega_y=2\pi\times9.5\times10^3$\,Hz, $\gamma_x=2\pi\times100\,$Hz, $\gamma_y=2\pi\times3\times10^{-3}\,$Hz, $T=4\times10^{-3}\,$K, $\mathcal E_y=2\times10^{4}\,\mathcal E_x=8\times10^{14}$\,Hz, $\kappa_x=10^3\kappa_y=9\times 10^8\,$Hz, $\omega_{c,y}=10^5\,\omega_{c,x}=2\pi\times3.7\times10^{15}\,$Hz.}
\end{figure}

{\it 2) Indirect measurement of non-classical correlation between cavity fields.} A viable strategy for the inference of the potentially non-classical nature of gravitational interaction goes through the assessment of possible non-classical correlations induced by the latter, according to the following rationale: The potential non-classical nature of gravity would induce a coupling between the $x$ and $y$ degrees of freedom, which might induce non-classical correlations in their joint state. Such a coupling disappears for classical gravity as $C^{(cl)}_2=0$. The induced all-mechanical correlations could in turn translate into analogous all-optical ones in light of the optomechanical coupling. In an experiment where all other plausible sources of correlations are carefully characterised, the possibility to detect all-optical quantum correlations would pave the way to the inference of the non-classical nature of gravity. It is important to stress that such correlations do not need to be as strong as entanglement: any non-zero value of $C^{(\gamma)}_2$ results in non-diagonal elements in the covariance matrix of the overall optomechanical system. The entries of such matrix are $\sigma_{ij}=\langle\{\delta \hat O_i,\delta \hat O_j\}\rangle$, where the expectation value is taken over the state of the system. Within the validity of the first-order expansion in the fluctuations invoked before, the presence of such non-diagonal elements entails non-classical correlations of the discord form~\cite{rmp}. It is thus sufficient to ascertain the non-nullity of the non-diagonal entries of the covariance matrix of the {\it all-optical} system embodied by the cavity fields only to infer, indirectly, the non classical nature of their correlations, and thus the quantum nature of the gravitational interaction. 

In Fig.~\ref{resMauro} {\bf (a)} we report the total norm $\sigma_{tot}=\sum_j |\sigma^{f}_{jj}|$ of the non-diagonal part of the covariance matrix $\sigma^{f}$ of the two cavity fields (i.e. we take only the fluctuation operators $\delta\hat O_i$ pertaining to the cavity fields) against $C_{1,x}$ for parameters such that $C_{1,x}=C_{1,y}$. We observe a linear growth of the covariances with the strength of the gravity-induced interaction. This gives rise to non-zero values of the discord between such fields, a illustrated in panel {\bf (b)}. Needless to say, the experimental ascertainment of a non-zero value of all-optical discord would pose significant experimental challenges, in light of its weakness. Nevertheless, the link with the strength of the non-diagonal entries of the corresponding covariance matrix offers a potentially viable route towards the goal of this paper: the reconstruction of the entries of an all-optical covariance matrix can indeed be accurately performed via high-efficiency homodyne measurements, as routinely implemented in many laboratories.

\begin{figure}[!]
{\bf (a)}\\
 \includegraphics[width=\columnwidth]{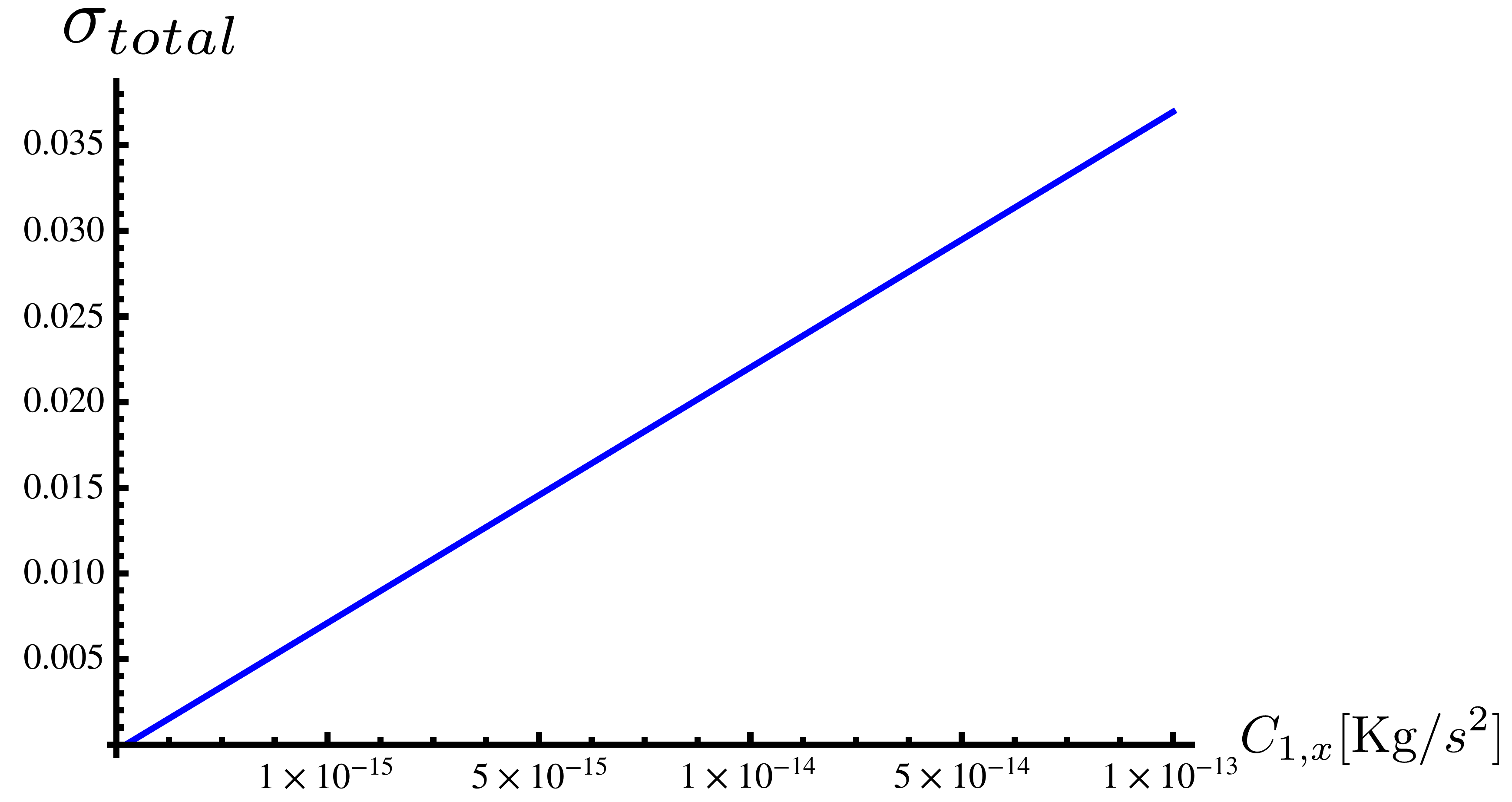}\\
 {\bf (b)}\\
 \includegraphics[width=\columnwidth]{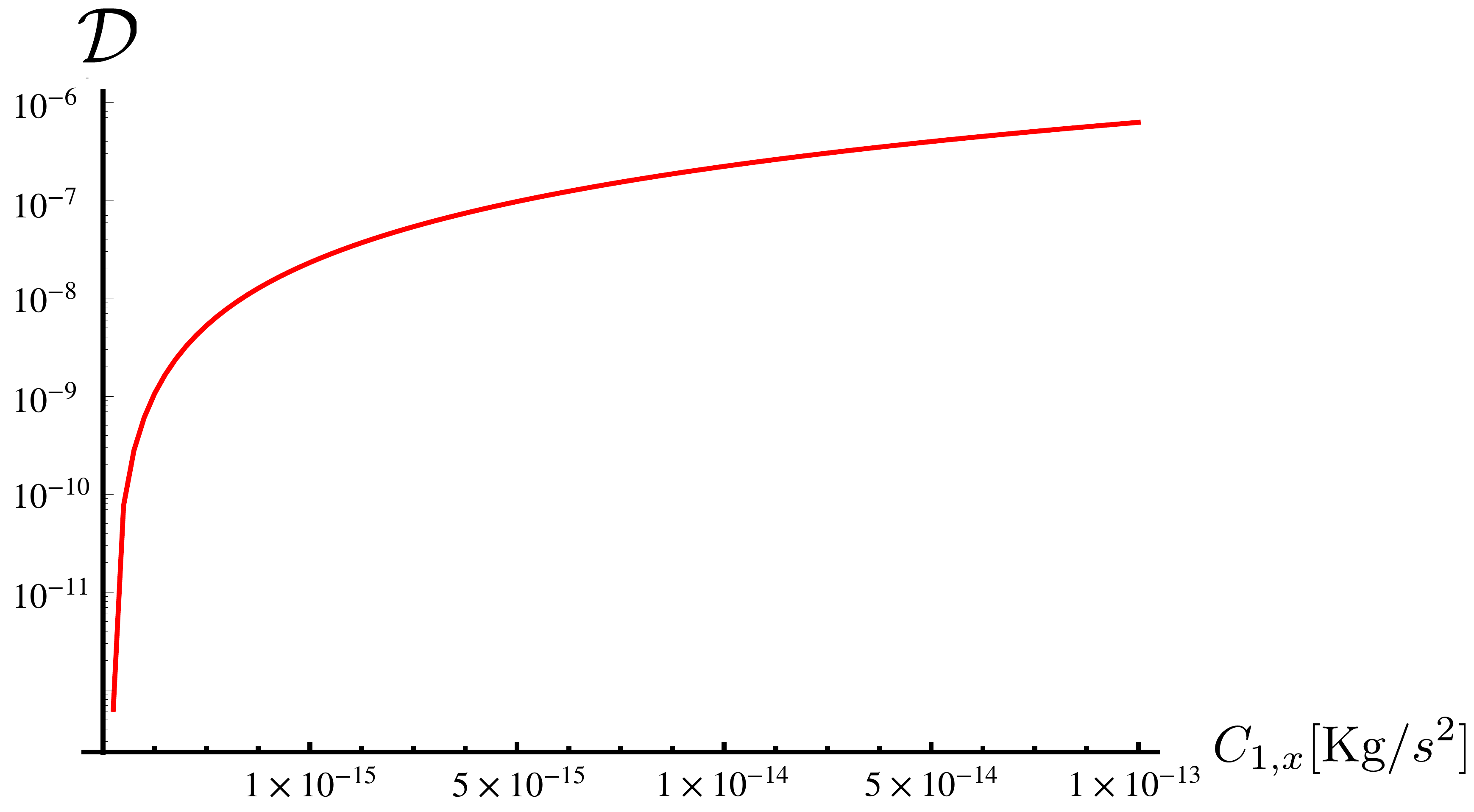}
 \caption{Total norm of the non-diagonal entries of the all-optical covariance matrix [panel {\bf (a)}] and all-optical discord [panel {\bf (b)}] plotted against $C_{1,x}$. We have taken $d_{x,y}\sim10^{-6}\,$m, $m_{1,2}=5\times 10^{-10}\,$Kg, mechanical modes of frequency $2\pi\times10^7\,$Hz, $T=4\,$mK, $\gamma_{x,y}=2\pi\times 100\,$Hz. The cavity has length of 1mm and finesse of $1.07\times10^4$. }
 \label{resMauro}
\end{figure}

{\it 3) Experimental feasibility.} To reduce the decoherence rates from gas collisions and blackbody photons to be smaller than the expected gravity effects, experiments should be done a low temperature and ultra-high vacuum. The calculation of the expected non-classical correlations quantified by discord have been done with typical parameters for optomechanical cantilever or membrane systems~\cite{Aspelmeyer}. The calculation for the direct observation of the DNS assumes parameters typical for levitated mechanical systems~\cite{Asenbaum2013, Kiesel2013, Millen2016, Winbdey2019, Delic2019,Jain,Vovrosh,Fonseca,Slezak}. The challenge for the direct DNS test will be to realise the strongly asymmetric double-cavity setup, where the two cavity frequencies are different.
 
The biggest challenge for the presented experimental geometry will be the handling of the effect of short-range interactions such as van der Waals \cite{Israel} and Casimir-Polder (CP)~\cite{Casimir1948}, which can overtone the gravity interaction between the two masses - given their close proximity. 

\section{Conclusions}
\label{conclusioni}
We have illustrated the dynamics of an optomechanical system probing the gravitational field of a massive quantum system in a spatial superposition. Two different dynamics are found whether gravity is treated quantum mechanically or classically. Here, we propose two distinct methods to infer which of the two dynamics rules the motion of the quantum probe, thus discerning the intrinsic \textit{nature} of the gravitational field. Such methods will be then eventually able to falsify one of the two treatments of gravity.

Recently other interferometric \cite{Bose} and non-interferometric \cite{Miao} tests of nature of gravity were proposed. They are based on the detection of entanglement between two probes, respectively coupled to two different massive systems, which interact through gravity (NV center spins for \cite{Bose} and cavity fields for \cite{Miao}). Clearly, to have such entanglement, each of the three couples of interconnected systems (probe 1, system 1, system 2 and probe 2) needs to be entangled on their own. Moreover, the entanglement between the two massive systems is inevitably small due to its gravitational character.
Conversely, our proposal profits of having only a single massive system involved in the interconnection, which reduces correlation losses. In addition, we provide a second method for discerning the nature of gravity: the individuation of a second peak in the DNS. The latter does not rely on delicate measurements of quantum correlations but can be assessed through standard optomechanical detection schemes. Other experimental proposals were presented in \cite{Mari,Anastopoulos,Belenchia,Hossenfelder}.

\noindent
{\it \bf Acknowledgements}.-- 
The authors acknowledge financial support from the H2020 FET Project TEQ (grant n.~766900) and the COST Action QTSpace (CA15220). AB acknowledges financial support from INFN.  MP is supported by the SFI-DfE Investigator Programme through project QuNaNet (grant 15/IA/2864), the Leverhulme Trust through the Research Project Grant UltraQuTe (grant nr.~RGP-2018-266) and the Royal Society Wolfson Fellowship scheme through project ExTraQCT (RSWF\textbackslash R3\textbackslash183013).



\end{document}